
\input epsf
\input harvmac

\noblackbox
\def\Title#1#2{\rightline{#1}\ifx\answ\bigans\nopagenumbers\pageno0\vskip1in
\else\pageno1\vskip.8in\fi \centerline{\titlefont #2}\vskip .5in}

%
%
\def\ajou#1&#2(#3){\ \sl#1\bf#2\rm(19#3)}

\font\ticp=cmcsc10

\def\grad{\nabla}
\def\dee{\partial}

%
%
\lref\suss{L. Susskind,``Some Speculations About Black Hole
Entropy in String Theory,''hep-th 9309145. }
\lref\sussug{L. Susskind and J. Uglum,``Black Hole Entropy in Canonical
Quantum Gravity and Superstring Theory,''hep-th 9401070.  }
\lref\waldjmp{J. Lee and R M. Wald, \ajou J. Math. Phys. &31 (90) 725.}
\lref\wald{R.M. Wald,``Black Hole Entropy is Noether Charge,''
 \ajou Phys. Rev. &D48 (93) 3427.}
\lref\iyer{V. Iyer and R.M. Wald,``Some Properties of Noether Charge
and a Proposal for Dynamical Black Hole Entropy,'' gr-qc 9403028. }
\lref\btz{M. Banados, C. Teitelboim, and J. Zanelli,
 ``Black Hole Entropy and the Dimensional Continuation of the
  Gauss-Bonnet Theorem,''\ajou Phys. Rev. Lett. &72 (94) 957.}
\lref\tseyt{A.A. Tseytlin,``Mobius Infinity Subtraction and Effective
Action in the $\sigma$-Model Approach to Closed String Theory,''
 \ajou Phys. Lett. &B208 (88) 221.}
\lref\jac{T. Jacobsen, G. Kang, and R. Myers,``On Black Hole Entropy,''
gr-qc 9312023. }

\Title{\vbox{\baselineskip12pt\hbox{UCSBTH-93-41}
\hbox{hepth@xxx/9406011}
}}
{\vbox{\centerline {A Comment on Black Hole}\vskip2pt
\centerline{ Entropy in String Theory}
}}

\centerline{{\ticp William  Nelson}\footnote{$^\dagger$}
{Email address: wnelson@sbphy.ucsb.edu}}
\vskip.1in
\centerline{\sl Department of Physics}
\centerline{\sl University of California}
\centerline{\sl Santa Barbara, CA 93106-9530}
\bigskip
\centerline{\bf Abstract}

In this note, we extend the string theoretic calculation of the black
hole entropy, first performed by Susskind and Uglum,
 away from the infinite mass limit.
It is shown that  the result  agrees with that obtained from the classical
action of string theory, using the Noether charge method
developed by Wald.
Also shown in the process is the equivalence of two general
techniques for finding black hole entropies-the Noether charge
method, and the
method of conical singularities.

\Date{6/94}

\newsec{Black Hole Entropy from String Theory}

One of the things we hope to get from a quantum theory of gravity is
a microscopic understanding of black hole thermodynamics. In particular
we'd like to see a microscopic structure associated with the horizon,
the number of whose states is counted by the usual black hole
entropy. Recently, it has been argued by
 Susskind \refs{\suss}, and Susskind and
Uglum \refs{\sussug}, that string theory may be able to provide this.

In particular, it was observed in \refs{\suss} that
 the string partition function
contains contributions which describe strings stuck onto the
horizon at their two endpoints; this stringy ``hair'' then
seems like a natural candidate for  microscopic structure.
Then in \refs{\sussug}, the authors made this idea more precise
by calculating the genus zero
contribution to the partition function, in the infinite mass limit,
and reproducing the expected result of $1\over 4$ per unit area.
Unfortunately these calculations are frought with peril, as they
require elements of off-shell string theory; we just have to use
the best available ansatz and hope. Perhaps the reasonable nature of
the results adds to its credibility.

In this paper, we extend the computation of \refs{\sussug}
away from the limit of infinite mass. Since we are
computing the classical (genus zero) part of the partition function,
we expect that the
answer should be the same as that obtained from the classical
string action, where the latter can be found using the Noether
charge technique developed by Wald \refs{\wald}.

The ansatz used in \refs{\sussug}, which we will also use here, is that
of Tseytlin \refs{\tseyt}. He argued that the string partition function and
the string action should be  closely related; indeed, he stated
\eqn\tsey{I={\dee\over\dee t}Z_R,}
where $Z_R$ is the renormalized genus zero sigma model partition function,
$t$ is the renormalization parameter, and $I$ is the classical string action,
alternately derivable from conformal invariance, or from scattering amplitudes.
(We use $I$ since $S$ will denote entropy.)
$Z_R$ still contains the (renormalized) Mobius infinity, and taking
$\dee\over\dee t$ is the prescription Tseytlin found for removing it.
Then the right hand side of \tsey\ should be identified with the genus zero
contribution to the generating functional $W$ (the quantity
usually defined in field theory
by $W=-\ln Z$), and we get $W=I$. From $W$ one gets the Helmholtz free
energy by $W=\beta F$.

Then $F$ determines the thermodynamics. In particular, the entropy is
given by the standard formula
\eqn\sstan{S=\beta^2 {\dee\over\dee\beta}F,}
where $\beta$ is the inverse temperature, proportional to the periodicity
of the regular euclidean continuation of the black hole spacetime.
The derivatives $\dee\over\dee\beta$ in \sstan\ direct us to vary this
periodicity, which creates a conical singularity in the spacetime.
We can then rewrite \sstan\ using $\epsilon$, the angular excess, instead of
$\beta$, getting
\eqn\s{S=(2\pi+\epsilon)^2{\dee\over\dee\epsilon}{1\over2\pi+\epsilon}I
    \, \big|_{\epsilon=0}.}
So the computation boils down to computing the first variation of the
Euclidean action under the introduction of a conical singularity;
see \refs{\btz,\sussug} for more on this idea. The
computation will be carried out in the next section, where its
equivalence to the Noether charge method will be demonstrated.

\newsec{Computing the Entropy}

First we review Wald's Noether charge method
(for further details see \refs{\wald,\jac,\iyer}). The starting point is a
covariant Lagrangian $L$, written as a $d$ form, where $d$ is
the number of dimensions. Then one computes the  variation of $L$ under
a diffeomorphism generated by an arbitrary vector field $\zeta$.
This can always be written schematically as
\eqn\lvar{\delta L = E^i \delta\psi_i+d\theta (\delta\psi_i),}
where $\psi_i$ are  the fields,
$\delta\psi_i$ are their variations,  $E^i$ are the equations of motion,
and $\theta$ is some $d-1$ form depending on the $\delta\psi_i$.
The condition of covariance of $L$ is
\eqn\cov{\delta L=d(\zeta\cdot L),}
where $\zeta\cdot L$ means $\zeta$ contracted onto the first index of $L$.
Then for on-shell fields ($E=0$), the last two equations imply
\eqn\cl{d(\theta-\zeta\cdot L)=0,}
so that $J\equiv \theta-\zeta\cdot L$ is a closed form; $J$ is just the dual of
the expected conserved current. Ordinarily one would not expect $J$
to be exact as well, but here the fact that it is closed for {\it all}
$\zeta$ allows one to prove exactness \refs{\waldjmp}. So
one has
\eqn\ex{J=dQ} for some $d-2$ form $Q$, which depends on $\zeta$.
The final step is to specialize $\zeta$ to be the horizon Killing
field for the black hole,
normalized to give unit surface gravity;
 then the entropy is identified as
\eqn\went{S=2\pi\int_HQ,}
where $H$ is the horizon $d-2$ surface.

Now for the method of conical singularities. We start with a somewhat formal
computation, and later sketch how to make it rigorous.
For simplicity, we specialize to the spherically symmetric case, and we choose
coordinates such that the euclidean metric is
\eqn\met{ds^2=dr^2+f(r)^2d\phi^2+g(r)d\Omega^2,}
where $f(r)\sim r^2$ as $r\sim 0$, and $d\Omega^2$ represents the other
$d-2$ coordinates, all angular, which play no role in the computation.
The euclidean time coordinate is $\phi$,
 which has a $2\pi$ period for regularity at $r=0$.

To add a  conical singularity with angular excess $\epsilon$ requires the
metric variation
\eqn\var{\delta g_{\phi\phi}={\epsilon\over\pi}f^2.}
In order to cast the computation in a form similar to the above, we look
for a vector field $\zeta$ which generates this variation via the usual
formula for diffeomorphisms, $\delta g_{ab}=\grad_{(a}\zeta_{b)}.$ This $\zeta$
cannot be smooth, since the metric variation in question does not
result from a diffeomorphism. The vector field which does the trick is
\eqn\zet{\zeta^a={\epsilon\over 2\pi}\phi {\dee\over\dee\phi}^a,}
which is smooth everywhere except on a cut at $\phi=0$. Note that this
is just ${\epsilon\over 2\pi}\phi$ times the horizon Killing vector used
in the Noether charge computation (in particular, $\dee\over\dee\phi$
is normalized for unit surface gravity.)

\topinsert
\centerline{\epsfysize= 3.0truein\epsfbox{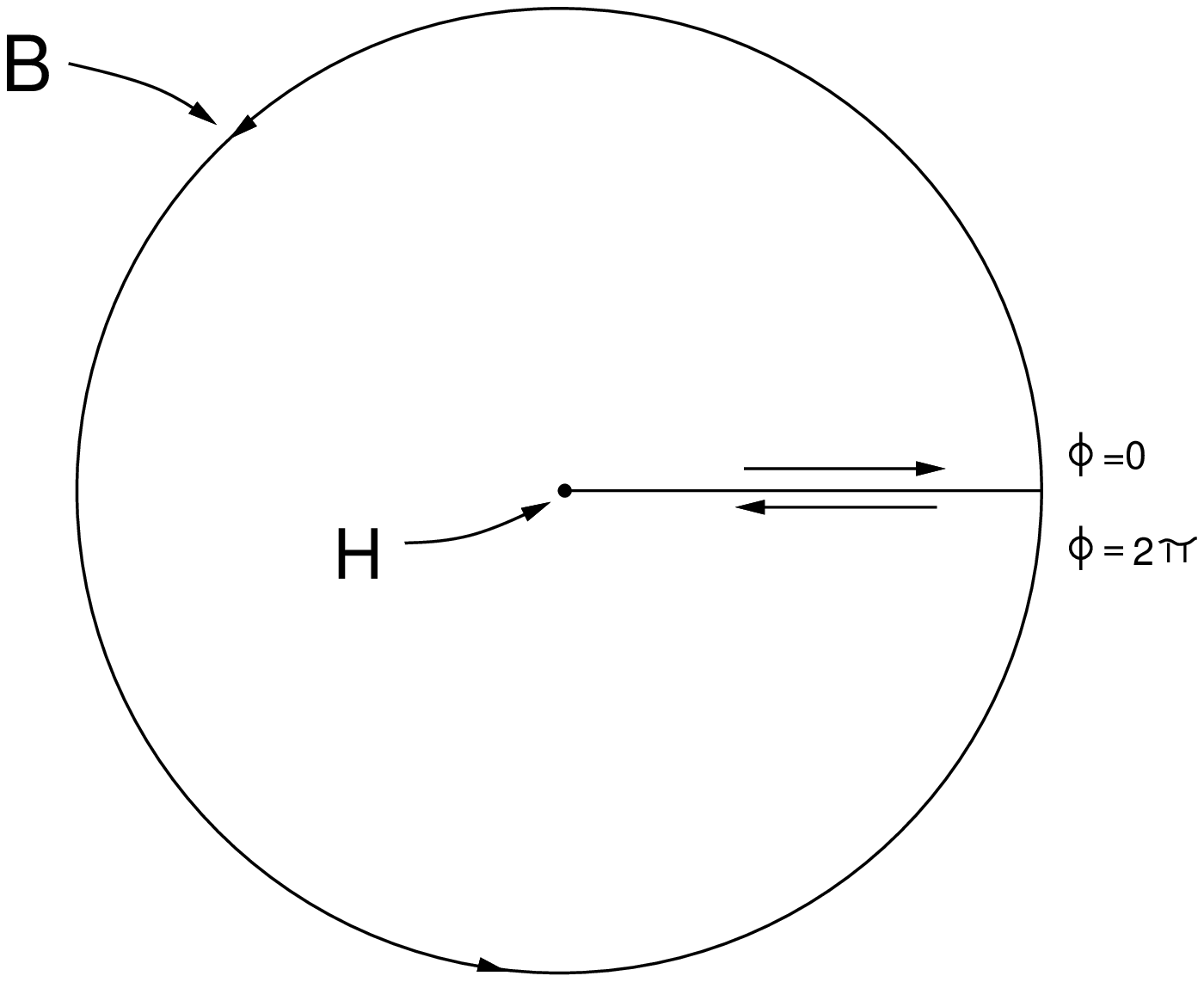}}
\noindent{{\bf Fig. 1}: $r,\phi$ cross section of the euclidean black
 hole spacetime, showing the cut at $\phi=0$ and the integration path B.}\par
\endinsert

{}From here the computation is almost the same as before. We need to
compute $\delta I = \int\delta L$. We write $\delta L$ in the
form of \lvar\ above. We use the on-shell condidition  $E=0$, and
 integrate the $d\theta$ term onto the boundary $B$, which consists of
both sides of the
cut  plus the asymptotic surface (see Fig. 1).
Then
 \eqn\ditwo{\eqalign{\delta I &=\int_{B}\theta  \cr
  &=\int_{B}J
 +{\epsilon\over 2\pi}\int_{B}(\phi{\dee\over\dee\phi}\cdot L)
 \cr
 &=\int_H Q\big|^{2\pi}_0
+\epsilon\int_{\phi=0}({\dee\over\dee\phi}\cdot L) ,\cr}}
where  $J$ and $Q$ are as defined above.
In the final step, we first used $J=dQ$ to integrate $J$ onto the
boundary of $B$, which we take to be $H|^{2\pi}_{0}$. Then we observed
that $\phi{\dee\over\dee\phi}\cdot L$ has no projection into the asymptotic
part of $B$, so its only contribution comes from the cut.

But whereas above $Q$ was evaluated for $\zeta={\dee\over\dee\phi}$, here
we have ${\epsilon\over 2\pi}Q(\phi{\dee\over\dee\phi})$.
It seems $\phi$ can't be factored out, since $\grad\phi$ terms may
appear-but what saves us is that
$\grad\phi$ is smooth
 across the cut, so that all $\grad\phi$ terms vanish from
$Q|^{2\pi}_0$. So we {\it can}  factor out the ${\epsilon\over
2\pi}\phi$, giving
\eqn\dithr{I+\delta I = \epsilon\int_H Q({\dee\over\dee\phi})
 +(2\pi +\epsilon)\int_{\phi=0}({{\dee\over\dee\phi}\cdot L}).}
(Here we also  used the $\dee\over\dee\phi$ symmetry to write
$I=2\pi\int_{\phi=0}{\dee\over\dee\phi}\cdot L$.)
Finally we compute $S$ by plugging into \s\ .
 Note that the second term is the classical
contribution of the fields in the spacetime away from the horizon; we expect
this to make no contribution to the entropy, and it doesn't, since it is
proportional to $2\pi+\epsilon$. The remainder gives the same result
obtained above, namely
$$S=2\pi\int_H Q,$$ with $Q$ evaluated on $\dee\over\dee\phi$.

Unfortunately, the above calculation is not rigorous, since for one thing,
 relevant
quantities such as $\nabla_a\phi$ are not defined at $r=0$, and for
another, the action $I$ will typically diverge if a conical singularity
is added (for example $\int R^2$ will diverge).
Here we outline a more rigorous path to the same conclusion.

\topinsert
\centerline{\epsfysize=4.0truein\epsfbox{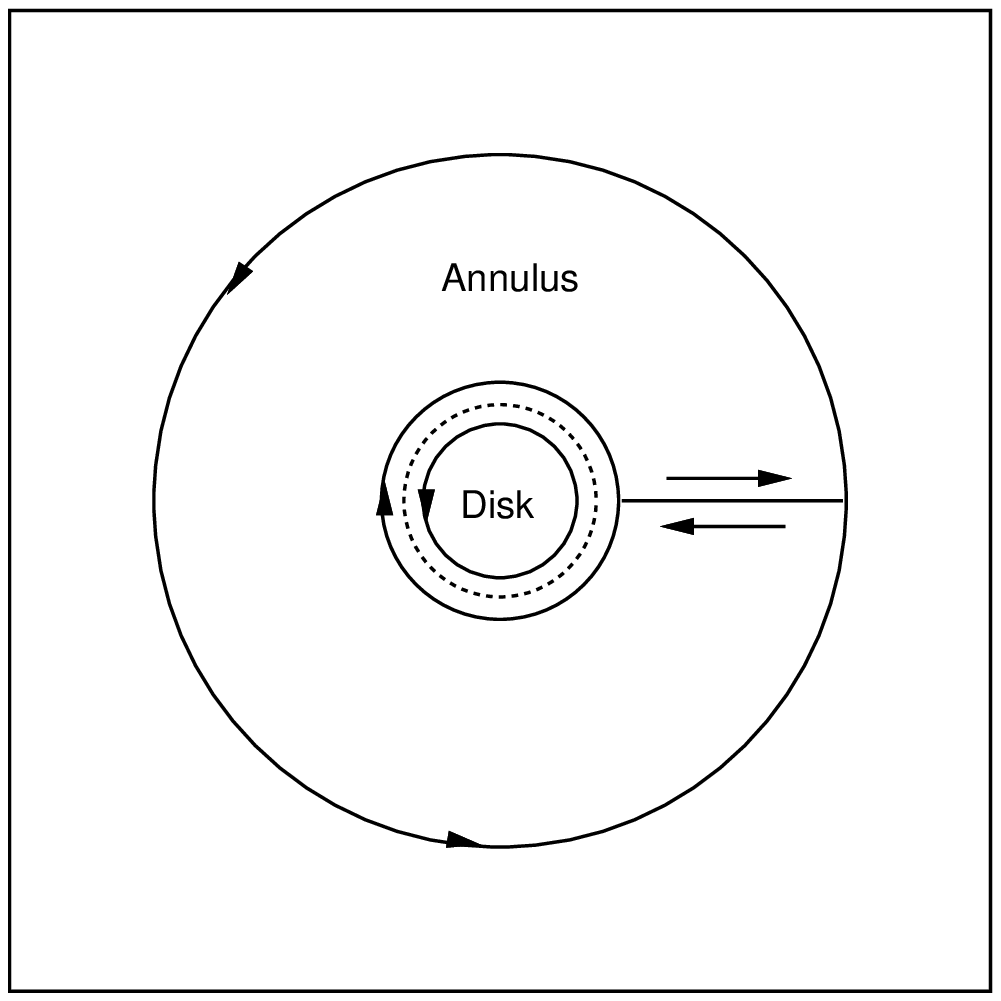}}
\noindent{{\bf Fig. 2}: The black hole spacetime with disk
 and annulus marked (the dotted line is the boundary.)
 Also shown are the paths of the boundary integrations.}
 \par
\endinsert

Starting from the euclidean black hole spacetime, we cut out a small
disk around $r=0$. Then, in the remaining annulus
(which extends to $r=\infty$), we choose again the metric variation generated
by $\zeta^a$ (eqn. \zet\ ). This gives the annulus a conical geometry
with angular excess $\epsilon$. Then, we choose some smooth metric variation
on the disk which matches smoothly onto that of the annulus. Then
we calculate the variation in the action due to these metric variations,
using \lvar. \footnote{$^\dagger$}{Note that \lvar\ holds for arbitrary
variations, although it was used above only for variations resulting
from a diffeomorphism.}
Finally, we
integrate $d\theta$ onto the boundaries, as in \ditwo.

 Now
there are two extra boundary segments, the inner boundary of the annulus,
and the outer boundary of the disk (see Fig 2). But since the metric
variation is smooth across the disk-annulus boundary, these extra
contributions simply cancel each other. The remaining path is just
$B$ from Fig. 1, except that the cut only extends to the disk boundary,
so in particular $r=0$ is avoided.
The final step is to take the limit as the disk shrinks
to zero radius, recovering the result \dithr\ above.

\bigskip\bigskip
\centerline{{\bf Acknowledgments}}\nobreak
I would like to thank
John Uglum and Joe Polchinski for helpful conversations, and Steve Giddings
and Rob Myers for
comments on earlier drafts.  This work was supported in
part by the grants DOE-91ER40618 and NSF PYI-9157463.

\listrefs

\end